\documentclass[%
 aip,
 amsmath,amssymb,
 reprint,%
]{revtex4-1}

\usepackage{graphicx}
\usepackage{dcolumn}
\usepackage{bm}

\usepackage[utf8]{inputenc}
\usepackage[T1]{fontenc}
\usepackage{mathptmx}
\usepackage{hyperref}
\usepackage{comment}
\usepackage[caption=false]{subfig}
\usepackage{color}

\begin{document}

\preprint{AIP/123-QED}

\title[The effect of heterogeneity on hypergraph contagion models]{The effect of heterogeneity on hypergraph contagion models}
\author{Nicholas W. Landry}
  \email{nicholas.landry@colorado.edu.}
\affiliation{Department of Applied Mathematics, University of Colorado at Boulder, Boulder, Colorado 80309}%
\author{Juan G. Restrepo}%
 \email{juanga@colorado.edu}
\affiliation{Department of Applied Mathematics, University of Colorado at Boulder, Boulder, Colorado 80309}%

\date{\today}

\begin{abstract}

The dynamics of network social contagion processes such as opinion formation and epidemic spreading are often mediated by interactions between multiple nodes. Previous results have shown that these higher-order interactions can profoundly modify the dynamics of contagion processes, resulting in bistability, hysteresis, and explosive transitions. In this paper, we present and analyze a hyperdegree-based mean-field description of the dynamics of the SIS model on hypergraphs, i.e. networks with higher-order interactions, and illustrate its applicability with the example of a hypergraph where contagion is mediated by both links (pairwise interactions) and triangles (three-way interactions). We consider various models for the organization of link and triangle structure, and different mechanisms of higher-order contagion and healing. We find that explosive transitions can be suppressed by heterogeneity in the link degree distribution, when links and triangles are chosen independently, or when link and triangle connections are positively correlated when compared to the uncorrelated case. We verify these results with microscopic simulations of the contagion process and with analytic predictions derived from the mean-field model. Our results show that the structure of higher-order interactions can have important effects on contagion processes on hypergraphs.

\end{abstract}

\maketitle

\begin{quotation}
Including group interactions in network models of contagion can significantly affect epidemic behavior. By studying the susceptible-infected-susceptible epidemic model on networks with higher-order interactions, we observe that for certain parameters there is a bistable regime, where above a critical number of infected individuals the contagion spreads until it becomes an epidemic, and below this critical number the epidemic dies out. We find that heterogeneity in the individual and group contact structure of a social network determines the existence of such tipping point events, and derive conditions for their appearance. Lastly, we comment on how three group contagion mechanisms -- collective contagion, infection by individuals, and the ``hipster effect'' -- affect the onset of epidemics and the existence of bistability.
\end{quotation}

\section{\label{sec:introduction}Introduction} 

The study of contagion processes is a fundamental problem in network science, with applications including epidemics\cite{pastor2015epidemic,trapman2007analytical,kiss2017mathematics,house2012modelling,newman2002spread,pastor2002epidemic,wang2003epidemic}, social media\cite{xiong2014opinion}, opinion formation\cite{watts2007influentials}, idea diffusion\cite{cowan2004network,valente1995network}, sudden changes in social convention\cite{gladwell2006tipping,centola2018experimental}, and many more. Contagion processes can be of many types, ranging from discrete-state models such as the susceptible-infected-susceptible (SIS) model, to continuous models of opinion formation, to realistic models of disease such as those currently used to model the spread of COVID-19 \cite{arenas2020mathematical,banerjee2020model}. Modeling the dynamics of such processes on pairwise interaction networks has been a hallmark of network science, providing many insights into the effect of network structure on the propagation of disease and information. Recently, the role of {\it complex contagion} mechanisms (i.e., contagion processes that can not be described solely by pairwise interactions) has received much attention \cite{battiston2020networks}. It has been shown that higher-order interactions in networks (i.e., interactions involving multiple nodes) can have profound effects on dynamical network processes\cite{porter2019nonlinearity+} such as opinion formation\cite{horstmeyer2020adaptive}, synchronization \cite{skardal2019abrupt,xu2020bifurcation,millan2020explosive} and population dynamics \cite{grilli2017higher}. Efforts to map higher-order interactions in real-world networks have uncovered rich structure \cite{benson2018simplicial} which is only now starting to be appreciated. In the context of contagion processes, it was recently shown \cite{iacopini2019simplicial} that the addition of higher-order interactions to the SIS epidemic model on Erd\"os-R\'enyi networks result in bistability, hysteresis, and explosive transitions to an endemic disease state (see also Refs.~\cite{de2020phase,cisernosmultigroup2020,de2020social,matamalas2020abrupt}).The simplicial SIS model has also been extended to scale-free uniform hypergraphs\cite{jhun2019simplicial}. The fact that the network SIS model with more general higher-order interactions results in bistability has been proven rigorously in Ref.~\cite{cisernosmultigroup2020}. However, so far there is no general theory explaining how heterogeneity and correlations in the structure of higher-order interactions affect the onset of bistability.

In this paper, we present and analyze a degree-based mean-field description of the dynamics of the SIS model in networks with higher-order interactions. To describe higher-order interactions we consider the SIS model on a {\it hypergraph}, formed by a set of nodes and a set of edges of multiple sizes (so that edges of size larger than two represent higher-order interactions). Our formulation allows us to consider heterogeneous structure in the organization of the edges of a given size, and correlations between the structure of edges of different sizes. Using the illustrative case of networks with edges of sizes 2 and 3, we derive conditions for the appearance of bistability and hysteresis in terms of moments of the degree distribution of the pairwise interaction network. We find that the onset of bistability and hysteresis can be suppressed by heterogeneity in the pairwise interaction network and promoted by positive correlations between the number of pairwise and higher-order interactions a node has. We also consider the effect of healing by higher-order interactions (a ``hipster effect'').

The structure of the paper is as follows. In Sec.~\ref{sec:model} we present our hypergraph and contagion models. In Sec.~\ref{sec:theory} we derive a mean-field description of the model and apply it to various illustrative cases. In Sec.~\ref{sec:critical_beta3} we study how model parameters affect the onset of bistability. Finally, we discuss our results and present our conclusions in Sec.~\ref{sec:discussion}.

\begin{figure}[b]
    \centering
    \includegraphics[width=6cm]{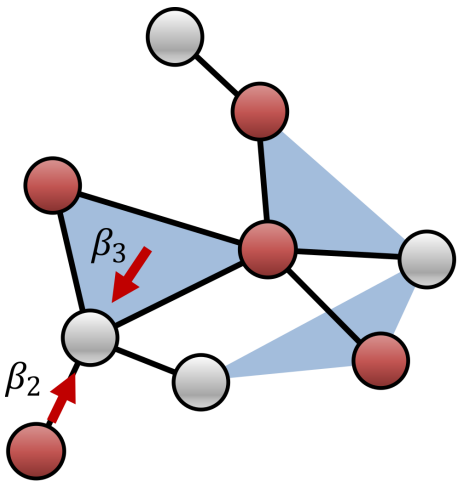}
    \caption{\label{fig:hypergraph}Illustration of a hypergraph. Infected nodes (red) infect a healthy node (grey) via hyperedges of sizes 2 and 3 with rates $\beta_2$ and $\beta_3$ respectively.}
\end{figure}

\section{\label{sec:model}Model}

In this section we present our hypergraph and contagion models. Our model consists of SIS contagion spreading on a hypergraph via pairwise and higher-order interactions. While we focus on the SIS epidemic model, we note that our formalism could be extended to other models. In the context of epidemic spreading, pairwise interactions could represent, for example, face-to-face interactions leading to contagion via viral droplets, while higher-order interactions could represent, for example, contagion via the shared spaces by a group. In the context of opinion dynamics, higher order contagion could model, for example, a majority-vote process common in caucusing. In the following, we provide details about the hypergraph model representing the higher-order interactions and the contagion models that we consider.

\subsection{\label{sec:model_hypergraph}Hypergraph model} 

We consider a population of $N$ nodes labeled $i = 1,2,\dots,N$ coupled via undirected hyperedges of sizes $m = 2,3,\dots,M$, where a hyperedge of size $m$ is a set of $m$ nodes, $\{i_1,i_2,\dots,i_m\}$. We define the {\it m-th order degree} of node $i$, $k_i^{(m)}$, as the number of hyperedges of size $m$ to which the node belongs, and its {\it hyperdegree} as the vector ${\bf k}_i = [k_i^{(2)},k_i^{(3)},\dots,k_i^{(M)}]$. The 2nd order degree of a node corresponds to the number of pairwise connections of the node, while higher-order degrees measure the node's participation in hyperedges of larger sizes. Figure \ref{fig:hypergraph} illustrates a hypergraph with hyperedges of sizes 2 and 3, which, for simplicity, we will henceforth denote as {\it links} and {\it triangles} respectively.

Extending degree-based descriptions of epidemic spreading on networks \cite{boguna2002epidemic,miller2012edge}, we will develop a mean-field theory for the propagation of epidemics based on the assumption that nodes with the same hyperdegree have the same statistical properties. For this purpose, we assume that the number of nodes with hyperdegree ${\bf k}$, $P({\bf k})$, is given, and that the probability that nodes with hyperdegrees ${\bf k}_1$, ${\bf k}_2$\dots, ${\bf k}_m$ belong to a hyperedge of size $m$ is given by $f_m({\bf k}_1, {\bf k}_2\dots, {\bf k}_m)$. This assumes that the statistical structure of the network is completely described by the hyperdegree distribution $P({\bf k})$ and the connection probabilities $f_m({\bf k}_1, {\bf k}_2\dots, {\bf k}_m)$. While this restriction rules out the possibility of assortative mixing by other node properties, it is straightforward to extend our formalism to include other node variables. Note that, counting the number of hyperedges of size $m$ in two different ways, the connection probabilities must be normalized such that
\begin{equation}
\frac{1}{m!}\sum_{{\bf k}_1,\dots, {\bf k}_m} P({\bf k}_1)\dots P({\bf k}_m)f_m({\bf k}_1, {\bf k}_2\dots, {\bf k}_m) = \frac{1}{m} \sum_{{\bf k}} k^{(m)}P({\bf k}).\label{eq:hyperedge_prob}
\end{equation}
For example, for the configuration model for networks without higher-order interactions (i.e., only hyperedges of size 2, $M = 2$), the hyperdegree of a node is just the number of links, ${\bf k} = k$, connecting that node to other nodes and the connection probability is $f_2(k,k') = k k'/(N\langle k \rangle)$, where $\langle k \rangle = \sum_{i=1}^N k_i/N = \sum_{k} k P(k)/N$. For networks with hyperedges of sizes 2 and 3, $f_3(k_1,k_2,k_3)$ is the probability that three nodes with degrees $k_1$, $k_2$, and $k_3$ are connected by a hyperedge of size 3. The configuration model for hypergraphs and its associated statistical properties has been studied in Refs.~\cite{courtney2016generalized,young2017construction}.

This framework allows us to study networks with heterogeneously distributed higher-order interactions and correlations between nodal degrees of different orders. In addition, it allows us to treat the case in which nodes belonging to a triangle are not necessarily connected by links, as is assumed in simplicial complex models\cite{iacopini2019simplicial}. We will study how the structure of higher-order interactions modifies some of the properties of epidemic spreading on networks with exclusively pairwise interactions (i.e., hyperedges of size 2 only), on which epidemic spreading has been studied extensively \cite{pastor2015epidemic}.

\begin{table}[b]
\begin{ruledtabular}
\begin{tabular}{p{2.5cm}p{4.5cm}}
\textbf{Variable} & \textbf{Definition}\\
\hline
$N$ & Number of nodes\\[0.05in]
$k^{(m)}$ & number of hyperedges of size $m$ a node belongs to\\[0.05in]
${\bf k} = [k^{(2)},\dots,k^{(M)}]$ & hyperdegree\\[0.05in]
$P({\bf k})$ & Number of nodes with hyperdegree ${\bf k}$\\[0.05in]
$\gamma$  & Rate of healing\\[0.05in]
$\beta_m$ & Rate of infection by a hyperedge of size $m$\\[0.05in]
$f_m({\bf k}_1, {\bf k}_2\dots, {\bf k}_m)$ & Probability that $m$ nodes form a hyperedge of size $m$\\[0.05in]
$x_{\bf k}$ & fraction of nodes with hyperdegree ${\bf k}$  that are infected\\[0.05in]
\end{tabular}
\end{ruledtabular}
\caption{\label{tab:notation}Relevant notation}
\end{table}

\subsection{\label{sec:model_contagion}Contagion model} 
Now we describe the contagion models we will study. As mentioned above, we will focus on the SIS model, but other epidemic models could be treated using the same formalism. We assume that at any given time $t\geq 0$, each node can be in either the susceptible (S) or infected (I) state. Infected nodes heal and become susceptible again at rate $\gamma$. Now we specify how hyperedges mediate the contagion process. In general, the probability of contagion by a hyperedge could be a function of the number of infected nodes in the hyperedge (e.g., as in Ref. \cite{de2020social}). Here we will consider the two extreme cases where contagion occurs if all the other members of the hyperedge are infected, or if at least one member of the hyperedge is infected. More precisely, in the {\it collective contagion} case, a susceptible node that belongs to a hyperedge of size $m$ gets infected at rate $\beta_m$ if \textit{all} the other members of the hyperedge are infected; in the {\it individual contagion} case the node gets infected  at rate $\beta_m$ if \textit{at least} one member is infected. While we will analyze these two cases only, in principle one could treat the case in which at least $j$ other nodes of the hyperedge need to be infected for contagion to occur using the techniques presented below. This case corresponds to a quorum of size $j$ and there is evidence for such effects in collective behavior\cite{ward2008quorum,pratt2002quorum}. For hyperedges of size $2$, i.e., links, both cases reduce to the usual contagion via pairwise interactions. The social contagion model of Ref.\cite{iacopini2019simplicial} corresponds to the collective contagion case. The contagion processes are illustrated in Figure \ref{fig:hypergraph} for hyperedges of sizes $2$ and $3$. Table \ref{tab:notation} summarizes the notation and variables used.

\section{\label{sec:theory}Mean-Field Analysis}

In this section we present a mean-field analysis of the epidemic dynamics on a network specified by the hyperdegree distribution $P({\bf k})/N$ and the hyperedge connection probabilities $f_m({\bf k}_1, {\bf k}_2\dots, {\bf k}_m)$. Assuming that all nodes with the same hyperdegree behave similarly, we focus on $x_{{\bf k}}$, the fraction of nodes with hyperdegree ${\bf k}$ that are infected. The mean-field equation describing the evolution of $x_{\bf k}$ is

\begin{widetext}
\begin{align}\label{eq:mf}
\frac{d x_{\bf k}}{d t}&=-\gamma x_{\bf k}+
(1-x_{\bf k})\sum\limits_{m=2}^M \beta_m \frac{1}{(m-1)!}\sum_{{\bf k}_1,\dots,{\bf k}_{m-1}} \prod\limits_{l=1}^{m-1} P({\bf k}_l) f_m({\bf k},{\bf k}_1,\dots,{\bf k}_{m-1})G(x_{{\bf k}_1},\dots,x_{{\bf k}_{m-1}})\\ 
&G(x_{{\bf k}_1},\dots,x_{{\bf k}_{m-1}}) =\left\{
\begin{array}{cc}
\prod\limits_{l=1}^{m-1}x_{{\bf k}_l}, & \text{collective contagion},\\
1-\prod\limits_{l=1}^{m-1}(1-x_{{\bf k}_l}), & \text{individual contagion}.
\end{array}\right.\label{eq:G}
\end{align}
\end{widetext}

The first term on the right-hand side of Eq.~\eqref{eq:mf} corresponds to healing at rate $\gamma$ and the second term accounts for infection by hyperedges. The number of hyperedges of size $m$ that can pass an infection to a node with hyperdegree ${\bf k}$  is calculated by considering all the possible hyperdegrees of the other $m-1$ nodes participating in the hyperedge (${\bf k},{\bf k}_1,\dots,{\bf k}_{m-1}$), counting how many such combinations there are not counting permutations [$ P({\bf k}_1)\cdots P({\bf k}_{m-1})/(m-1)!$], calculating what fraction of such combinations form a hyperedge with the node in consideration [$f_m(x_{\bf k},x_{{\bf k}_1},\dots,x_{{\bf k}_{m-1}})$], multiplying by the probability that the hyperedge can transmit the infection [$G(x_{{\bf k}_1}\cdots x_{{\bf k}_{m-1}})$], and summing over all hyperdegree combinations. The probability that the hyperedge can transmit the infection, given by (\ref{eq:G}), depends on whether the collective contagion or individual contagion model is assumed. Note that the form for $G$ taken above, and the mean field treatment in general, assume that the states of nodes are independent. A better approximation that includes correlations between connected nodes has been implemented in Ref.~\cite{matamalas2020abrupt} for the case of unstructured hyperedges of sizes $2$ and $3$, leading to improved quantitative agreement with the results of numerical simulations. Since our interest is in the effects of higher-order structures on qualitative aspects of the epidemic dynamics, we will use the mean field approximation in Eq.~(\ref{eq:mf}). A similar mean-field equation for a node-based description of the contagion process was recently formulated in Ref.~\cite{cisernosmultigroup2020}. In the following we will apply the mean-field description to illustrative cases.

\subsection{\label{sec:theory_2and3}Hyperedges of sizes 2 and 3 with collective contagion}

Here we focus on the case where the hyperedge sizes are either $2$ or $3$, i.e., $M = 3$. This corresponds to a network like in Fig. \ref{fig:hypergraph}, with hyperedges of size $2$ (links) and $3$ (triangles). For simplicity, here we denote the number of links per node as $k$, i.e., $k = k^{(2)}$, and the number of triangles a node belongs to by $q$, i.e., $q = k^{(3)}$. In addition, we will consider the case where the connection probabilities depend only on the node links, i.e., $f_m({\bf k},{\bf k}_1,\dots,{\bf k}_{m-1}) = f_m({k,k_1,\dots,k_{m-1}})$. With these assumptions, and using the collective contagion rule in Eq.~(\ref{eq:G}), Eq. (\ref{eq:mf}) becomes 
\begin{widetext}
\begin{align}\label{eq:mfreduced}
\frac{d x_{k,q}}{d t}&=-\gamma x_{k,q}+
(1-x_{k,q})\beta_2 \sum_{k_1,q_1} P(k_1,q_1)f_2(k,k_1)x_{k_1,q_1} +(1-x_{k,q})\frac{\beta_3}{2} \sum_{k_1,q_1,k_2,q_2} P(k_1,q_1)P(k_2,q_2)f_3(k,k_1,k_2)x_{k_1,q_1}x_{k_2,q_2},
\end{align}
\end{widetext}
where the first term on the right hand side represents healing, the second represents contagion by links, and the third represents contagion by triangles.

\begin{figure}[t]
    \centering
    \includegraphics[width=8cm]{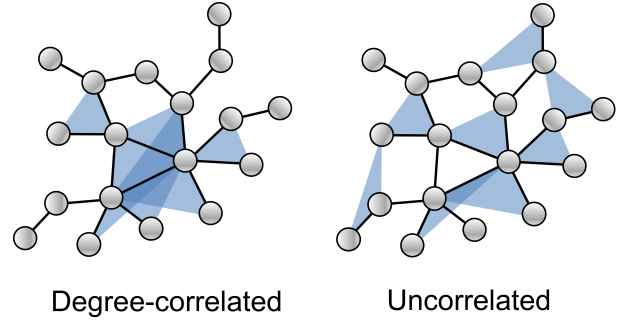}
    \caption{\label{fig:degree-correlated_vs_uncorrelated}Schematic illustration of the degree-correlated and uncorrelated cases. In the degree-correlated case (left), nodes with more links are more likely to belong to a triangle. In the uncorrelated case (right), triangles connect nodes with a probability independent of their degree.}
\end{figure}

Since the connection probabilities do not depend on $q$, we can reduce the dynamics to the fraction of nodes with degree $k$ that are infected, 
\begin{align}
x_k = \frac{\sum_q P(k,q) x_{k,q}}{P(k)},
\end{align}
where $P(k) = \sum_q P(k,q)$ is the number of nodes with degree $k$. Multiplying Eq.~(\ref{eq:mfreduced}) by $P(k,q)$, summing over $q$ and dividing by $P(k)$, we obtain
\begin{align}\label{eq:mfmarginal}
\frac{d x_k}{d t}&=-\gamma x_k+
(1-x_k)\beta_2 \sum_{k_1} P(k_1)f_2(k,k_1)x_{k_1}\\ &+(1-x_k)\frac{\beta_3}{2} \sum_{k_1,k_2} P(k_1)P(k_2)f_3(k,k_1,k_2)x_{k_1}x_{k_2}.\nonumber
\end{align}
For the link connection probability $f_2(k,k_1)$, we will take $f_2(k,k_1) = k k_1/(N\langle k \rangle)$, which corresponds to nodes being connected completely at random according to their degree as in the configuration model. For the triangle connection probability $f_3$, we will consider two cases: the {\it uncorrelated} case and the {\it degree-correlated case}. In the degree-correlated case, we assume that the connection probability is given by $f_3(k,k_1,k_2) = 2k k_1 k_2/(N\langle k \rangle)^2$, so that nodes that have a higher number of links also belong to more triangles. In the uncorrelated case, we assume instead that $f_3(k,k_1,k_2) = 2\langle k \rangle/N^2$, 
so that triangles are formed independent of the nodal degrees. The normalization is chosen using Eq.~(\ref{eq:hyperedge_prob}) so that the mean number of triangles per node, $\langle q\rangle=\sum_{i=1}^N k_i^{(3)}/N$, in each case is equal to $\langle k \rangle$. We note that the model for triangle formation in Ref.~\cite{iacopini2019simplicial} corresponds to the uncorrelated case. We can choose the mean triangle degree independent of the mean network degree by scaling $f_3(k,k_1,k_2)$ by $\langle q\rangle/\langle k\rangle$, but for simplicity, we assume $\langle q\rangle=\langle k\rangle$. Figure \ref{fig:degree-correlated_vs_uncorrelated} illustrates the difference between the two cases in a small network, where in the degree-correlated case, the triangles cluster around nodes with high pairwise degree, and in the uncorrelated case, the triangles are distributed uniformly at random on the network.

We can also specify the distribution of triangle degrees by defining $f_2(q,q_1)$ and $f_3(q,q_1,q_2)$ and then reducing Eq.~(\ref{eq:mfreduced}) by multiplying by $P(k,q)$, dividing by $P(q)$, and summing over $k$ to reduce the dynamics to the fraction of infected nodes with triangle degree $q$. For the triangle connection probability, we take $f_3(q,q_1,q_2)=2qq_1q_2/(N\langle q\rangle)^2$ which corresponds to three nodes being connected at random according to the configuration model for triangles\cite{courtney2016generalized}. For the pairwise links, we define the degree-correlated and uncorrelated cases as before, where in the degree-correlated case, $f_2(q,q_1)=qq_1/(N\langle q\rangle)$, and for the uncorrelated case, $f_2(q,q_1)=\langle q\rangle/N$. From there, we can use the same formalism as our approach when specifying the pairwise degree.

Now we consider separately the degree-correlated and uncorrelated cases. In the correlated case, where $f_3(k,k_1,k_2) = k k_1 k_2/(N\langle k \rangle)^2$, Eq.~(\ref{eq:mfmarginal}) can be rewritten in terms of the fraction of infected links
\begin{eqnarray}
V&&=\sum\limits_k \frac{k P(k) x_k}{N\langle k\rangle}\label{eq:vdef}
\end{eqnarray}
as
\begin{equation}
    \frac{d x_k}{d t}=-\gamma x_k+\beta_2(1-x_k)k V+\beta_3(1- x_k)k V^2.\label{eq:dyncorrelated}
\end{equation}

In this case, the dynamics of nodes of degree $k$ is determined by the global variable $V$. To study the qualitative characteristics of the dynamics, we find the steady-state solutions. The fixed point of Eq.~(\ref{eq:dyncorrelated}) is
\begin{equation}
x_k=\frac{\beta_2 k V +\beta_3 kV^2}{\gamma+\beta_2 kV +\beta_3 kV^2}.
\label{eqn:equilibrium_dc}
\end{equation}
Inserting this in (\ref{eq:vdef}), we obtain a nonlinear equation that determines the fraction of infected links $V$:
\begin{equation}
    V=\frac{1}{N\langle k \rangle}\sum\limits_k \frac{k P(k)(\beta_2 k V +\beta_3 k V^2)}{\gamma +\beta_2 k V +\beta_3 k V^2}.\label{eq:v1}
\end{equation}
The state with no infection, $V = 0$, is a solution to (\ref{eq:v1}). However, it is linearly unstable for  $\beta_2 > \beta_2^c = \gamma \langle k\rangle/\langle k^2\rangle$, as can be seen by linearizing Eq.~\eqref{eq:dyncorrelated} about $V = 0$, multiplying by $kP(k)/(N\langle k\rangle)$, and summing over $k$, which yields the linearized equation for the evolution of the perturbation $\delta V$
\begin{align}
 \frac{d \delta V}{d t}=-\gamma \delta V+\beta_2 \frac{\langle k^2\rangle}{\langle k\rangle} \delta V.
\end{align}
The nonzero solutions of Eq.~(\ref{eq:v1}) represent states with a nonzero fraction of infected nodes. 

Now we study the uncorrelated case where $f_3(k,k_1,k_2) = 2\langle k \rangle/N^2$. In this case, Eq.~(\ref{eq:mfmarginal}) can be rewritten in terms of the fraction of infected nodes 
\begin{equation}
U=\sum\limits_k \frac{P(k) x_k}{N},\label{eq:udef}
\end{equation}
and the fraction of infected links $V$. In terms of these quantities, Eq.~(\ref{eq:mfmarginal}) reads
\begin{equation}
    \frac{dx_k}{dt}=-\gamma x_k+\beta_2(1-x_k)kV+\beta_3(1- x_k)\langle k\rangle U^2.\label{eq:uv}
\end{equation}
As in the prior case, the equilibrium is
\begin{equation}
x_k=\frac{\beta_2 kV+\beta_3 \langle k\rangle U^2}{\gamma+\beta_2 kV+\beta_3 \langle k \rangle U^2}.
\label{eqn:equilibrium_uc}
\end{equation}
Evaluating this expression in Eqs.~(\ref{eq:vdef}) and (\ref{eq:udef}) we obtain the coupled equations
\begin{align}
    U&=\frac{1}{N}\sum\limits_k \frac{P(k)(\beta_2 kV+\beta_3 \langle k\rangle U^2)}{\gamma+\beta_2 kV+\beta_3 \langle k \rangle U^2},\label{eq:ueq}\\
    V&=\frac{1}{N\langle k\rangle}\sum\limits_k \frac{kP(k)(\beta_2 kV+\beta_3 \langle k\rangle U^2)}{\gamma+\beta_2 kV+\beta_3 \langle k \rangle U^2}.\label{eq:veq}
\end{align}

The state with no infection, $U=0$, $V = 0$, is a solution of \eqref{eq:ueq}-\eqref{eq:veq}. By considering perturbations $\delta U$, $\delta V$ from this solution, linearizing Eq.~(\ref{eq:uv}), and evaluating in Eq.~(\ref{eq:vdef}) for the first equation and Eq.~(\ref{eq:udef}) for the second equation, we obtain the linear system
\begin{align}
 \frac{d \delta V}{d t}=-\gamma \delta V+\beta_2 \frac{\langle k^2\rangle}{\langle k\rangle} \delta V,\\
 \frac{d \delta U}{d t}=-\gamma \delta U+\beta_2 \langle k\rangle \delta V,
\end{align}
which shows that the no infection state is linearly unstable for $\beta_2 > \gamma\langle k\rangle/\langle k^2\rangle$, which is the same threshold we obtained for the correlated case. 

\begin{figure}
    \centering
    \includegraphics[width=8cm]{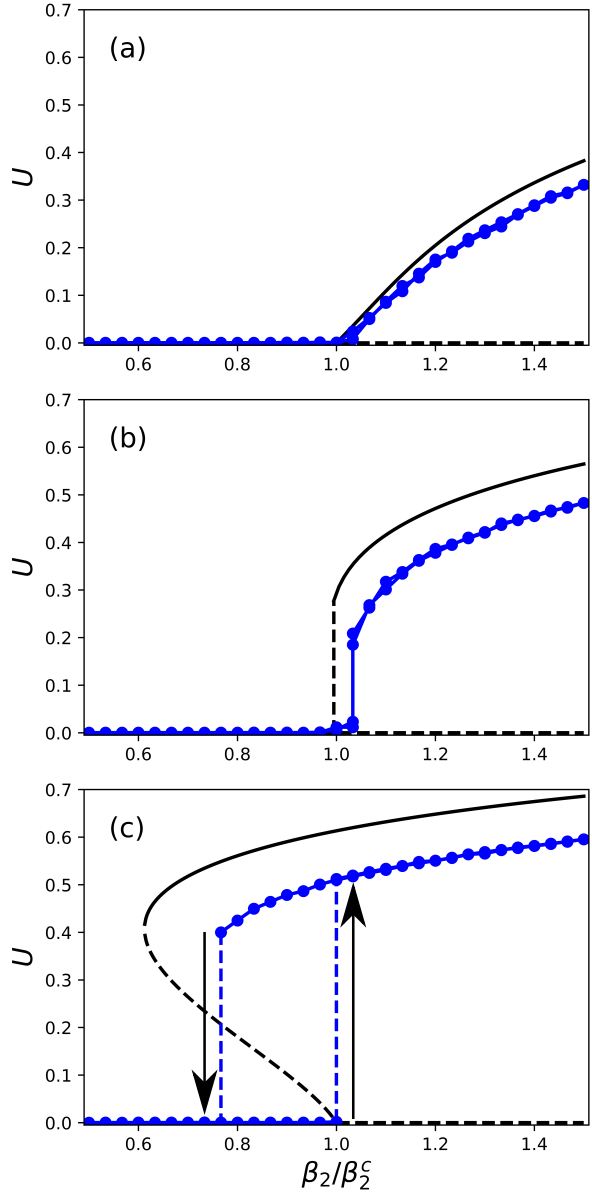}
    \caption{Fraction of infected nodes $U$ versus link infectivity $\beta_2$ obtained from the mean field equations  (\ref{eq:ueq})-(\ref{eq:veq}) (solid and dashed lines) and from microscopic simulations (connected circles) using $P(k)\propto k^{-4}$ on $[67,1000]$, $\gamma = 2$, and $N = 10000$ for $\beta_3 = 0.0194$ (a),  $0.0388$ (b), and $0.05482$ (c). Refer to the text for an explanation of the discrepancy between the mean field equations and microscopic simulations.}
    \label{fig:infection_mf}
\end{figure}

In summary, nonzero solutions of Eq.~(\ref{eq:v1}) and Eqs.~(\ref{eq:ueq})-(\ref{eq:veq}) for the degree-correlated and uncorrelated cases, respectively, represent states with a nonzero number of infected nodes. Figure \ref{fig:infection_mf} shows the fraction of infected nodes $U$ for the uncorrelated case as a function of the normalized pairwise infectivity $\beta_2/\beta_2^c$ for three values of the triangle infectivity $\beta_3$ obtained from numerical solution of Eqs.~(\ref{eq:ueq})-(\ref{eq:veq}) with $P(k) \propto k^{-4}$ for $67<k<1000$ and 0 otherwise. Different solutions are plotted as solid and dashed lines to indicate stability or instability, respectively.
The connected circles are obtained from numerical simulations of the full stochastic microscopic model. In these simulations $\beta_2$ was slowly increased in small steps up to a maximum value, and subsequently decreased back to its initial value. For each $\beta_2$, the average number of infected nodes after transient effects disappeared is shown as a filled circle. For more details about the simulations, see Appendix \ref{sec:appendix_numerics}.

The behavior of the microscopic simulations is captured qualitatively by the mean field equations. The quantitative disagreement is likely due to the assumptions inherent to the mean-field approximation. In fact, Ref.~\cite{matamalas2020abrupt} has shown that, for the particular case of uncorrelated triangles on an Erd\"os-R\'enyi network, the disagreement almost disappears when pair correlations are taken into account. Since our interest in this paper is on the qualitative dynamics, we use the mean-field theory, but note that the approaches proposed in \cite{matamalas2020abrupt,guerra2010annealed} could be used to obtain better approximations. The qualitative aspects of interest, captured by the mean field equations and the numerical solution of Eqs.~(\ref{eq:ueq})-(\ref{eq:veq}), are the following.
For small values of $\beta_3$ [Fig.~\ref{fig:infection_mf}(a), $\beta_3=0.0194$] the bifurcation from the state with no infection ($U = 0$) to the infected state ($U>0$) is continuous. However, for larger values of $\beta_3$ [Fig.~\ref{fig:infection_mf}(c), $\beta_3=0.0582$], the transition is discontinuous: as $\beta_2$ increases past a critical value $\beta_2^c$, the fraction of infected links increases explosively towards an epidemic equilibrium (upward arrow). If $\beta_2$ is subsequently decreased, the fraction of infected links remains high until $\beta_2$ decreases past the value at which the epidemic equilibrium solution disappears, and then it decreases to zero (downward arrow). For such values of $\beta_3$, there is hysteresis, bistability, and explosive transitions. At a critical value $\beta_3 = \beta_3^c$, which will be the focus of our interest, there is a transition from the type of bifurcation shown in Fig~\ref{fig:infection_mf}(a) to the type of bifurcation shown in Fig.~\ref{fig:infection_mf}(c). Fig.~\ref{fig:infection_mf}(b) shows $U$ as a function of $\beta_2$ for a value $\beta_3=0.0388\approx \beta_3^c$. We are interested in exploring how the presence of this bistable regime is affected by the degree distribution $P(k)$ and other parameters of the model, in particular the triangle infectivity, $\beta_3$.

\begin{figure}[t]
    \centering
    \includegraphics[width=8cm]{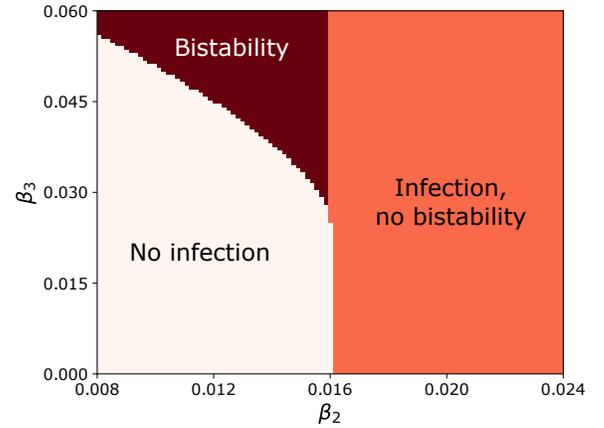}
    \caption{\label{fig:phase_space_collective}Phase diagram for the degree-correlated, collective contagion model. The light pink region labeled ``No infection'' corresponds to $1$ solution of Eq.~(\ref{eq:v1}), the orange region labeled ``Infection, no bistability'' to $2$ solutions, and the region labeled ``Bistability'' to $3$ solutions. The parameters are $\gamma = 2$ and $P(k) \propto k^{-4}$ when $67 < k < 1000$ and $0$ otherwise.}
\end{figure}

Figure~\ref{fig:phase_space_collective} shows the phase diagram in the $(\beta_2,\beta_3)$ plane for the degree-correlated, collective contagion model. The plot was obtained by counting the number of solutions of Eq.~(\ref{eq:v1}) as a function of $\beta_2$ and $\beta_3$ for $\gamma = 2$, and $P(k) \propto k^{-4}$ when $67 < k < 1000$ and $0$ otherwise (all subsequent phase diagram plots are calculated using the same parameters). Light pink indicates that there is only the solution $V = 0$ corresponding to a stable state with no contagion. Orange indicates two solutions, the unstable $V = 0$ solution and another stable solution with $V>0$. Finally, dark red indicates a bistable regime with three solutions: the stable $V = 0$ solution, and a pair of stable and unstable solutions with positive $V$. As noted in Ref.~\cite{iacopini2019simplicial,cisernosmultigroup2020}, this regime is only present for large enough triangle infectivity, i.e., for $\beta_3 > \beta_3^c$. The phase space for the uncorrelated case (not shown) is qualitatively similar to the one in Fig.~\ref{fig:phase_space_collective}, but the transition to bistable behavior occurs at a larger value of $\beta_3$.

\begin{figure}
\begin{center} 
\includegraphics[width=8cm]{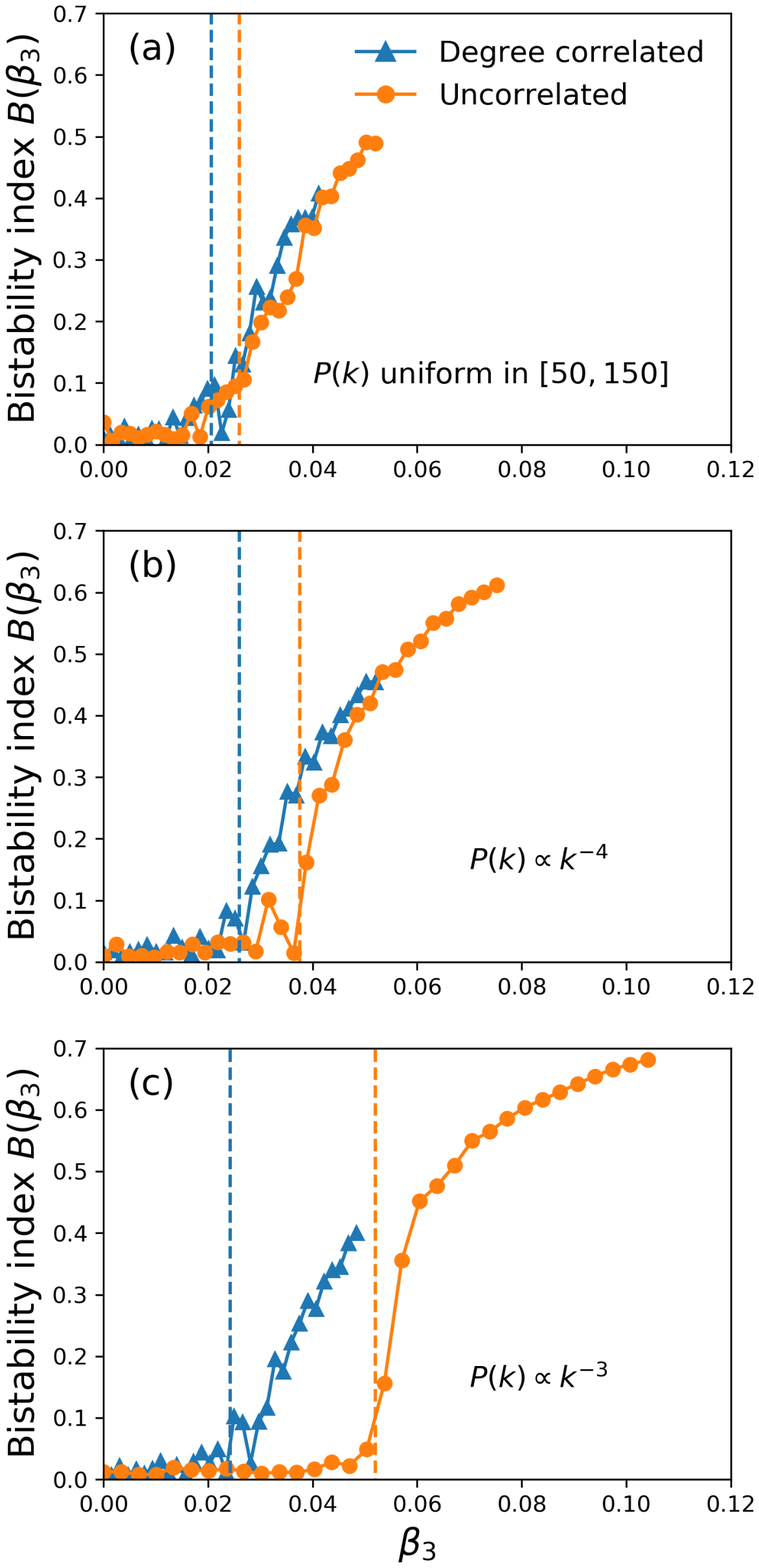}\\
    \caption{Bistability index $B$ as a function of $\beta_3$ for (a) $P(k)$ constant for $50 < k < 150$ and $0$ otherwise, (b) $P(k) \propto k^{-4}$ for $67 < k < 1000$ and $0$ otherwise, and (c)  $P(k) \propto k^{-3}$ for $53 < k < 1000$ and $0$ otherwise. For each distribution, we considered the uncorrelated case (orange connected circles) and the degree correlated case (blue connected triangles). The dashed lines indicate the value $\beta_3^c$ at which we expect the onset of bistability, obtained from the numerical solution of the mean field equations (\ref{eq:udef}) and (\ref{eq:ueq})-(\ref{eq:veq}).}
    \label{fig:bistability_index}
\end{center}
\end{figure}

To quantify how the onset of bistability depends on the hypergraph parameters, we define the {\it bistability index} $B(\beta_3)$ as the maximum separation, over all values of $\beta_2$, between the largest and smallest stable solutions for the fraction of infected nodes $U$. The bistability index can be calculated from microscopic simulations of the contagion process such as those used to produce Fig.~\ref{fig:infection_mf}, or from numerical solution of Eq.~(\ref{eq:v1}) for the correlated case and  Eqs.~(\ref{eq:ueq})-(\ref{eq:veq}) for the uncorrelated case.
In Figure~\ref{fig:bistability_index} we plot the bistability index $B$ as a function of $\beta_3$ computed from microscopic simulations for three choices of the link degree distribution $P(k)$, all with a mean degree of 100: (a) $P(k)$ constant for $50 < k < 150$ and $0$ otherwise, (b) $P(k) \propto k^{-4}$ for $67 < k < 1000$ and $0$ otherwise, and (c)  $P(k) \propto k^{-3}$ for $53 < k < 1000$ and $0$ otherwise. For each distribution, we considered the uncorrelated case (orange connected circles) and the degree correlated case (blue connected triangles). The dashed lines indicate the value $\beta_3^c$ at which we expect the onset of bistability, obtained from numerical solution of Eqs.~(\ref{eq:v1}) and (\ref{eq:ueq})-(\ref{eq:veq}) for the degree correlated and uncorrelated cases, respectively (in Sec.~\ref{sec:critical_beta3} we provide analytical expressions for these values). As the degree distribution of the pairwise interaction network $P(k)$ becomes more heterogeneous from (a) to (c), the value of $\beta_3$ at which the onset of bistability occurs increases for the uncorrelated case, while it remains almost unchanged for the degree-correlated case. A heuristic interpretation of this phenomenon is the following: in the uncorrelated case, the triangle interactions do not depend on the heterogeneity of the link degree distribution. Therefore, as the link degree distribution $P(k)$ becomes more heterogeneous, contagion becomes dominated by hubs of the pairwise interaction network, a mechanism which does not result in bistability. Therefore, bistability is suppressed in the uncorrelated case. On the other hand, for the degree correlated case, both triangle and link contagion mechanisms increase their effectiveness in tandem as the heterogeneity of the link degree distribution is increased. It is important to note that the increase in $\beta_3^c$ with heterogeneity, which is shown here in absolute terms, still occurs if one considers it relative to the value of $\beta_2^c$ (i.e., $\beta_3^c/\beta_2^c$ also increases with heterogeneity), as we will show later.

Another interesting aspect seen in Figure~\ref{fig:bistability_index} is that the transition to bistable behavior seems sharper in the uncorrelated case for the more heterogeneous networks. As we will see in Sec.~\ref{sec:critical_beta3}, the nature of the bifurcation is indeed different for the uncorrelated case and heterogeneous networks. 

Finally, we have to point out that the numerical calculation of the bistability index from numerical simulations can be challenging. When the unstable solution is small, finite size effects can cause transitions to the nonzero stable solution from the stable zero solution, making the numerical determination of the stable fixed points difficult and the bistability index plots noisy. Nevertheless, the mean field theory predicts well the onset of bistability.

\subsection{\label{sec:theory_individual}Hyperedges of sizes 2 and 3 with individual contagion}

Now we consider the case of individual contagion, in which an $m$-hyperedge infects a susceptible node with rate $\beta_m$ when {\it at least} one member of the hyperedge is infected. For simplicity, we will still consider only links and triangles ($M=3$) with infection rates of $\beta_2$ and $\beta_3$ respectively.

The analogue to Eq.~(\ref{eq:mfmarginal}) for the individual contagion case is
\begin{align}\label{eq:mfmarginal2}
&\frac{d x_k}{d t}=-\gamma x_k+
(1-x_k)\beta_2 \sum_{k_1} P(k_1)f_2(k,k_1)x_{k_1}+\\ 
&(1-x_k)\frac{\beta_3}{2} \sum_{k_1,k_2} P(k_1)P(k_2)f_3(k,k_1,k_2)[1 -(1-x_{k_1})(1-x_{k_2})].\nonumber
\end{align}
For the correlated case, $f_3(k,k_1,k_2) = 2k k_1 k_2/(N\langle k \rangle)^2$, this can be rewritten as
\begin{equation}
    \frac{dx_k}{dt}=-\gamma x_k+(\beta_2+2\beta_3)(1-x_k)kV - \beta_3(1-x_k)kV^2,
\end{equation}
with fixed point
\begin{equation}
    x_k=\frac{(\beta_2 + 2\beta_3) kV - \beta_3 kV^2}{\gamma+(\beta_2 + 2\beta_3) kV - \beta_3 kV^2}.
\end{equation}
Inserting this into Eq.~(\ref{eq:vdef}) like before, we obtain
\begin{equation}
    V=\frac{1}{N\langle k\rangle}\sum\limits_{k} \frac{k P(k)[(\beta_2+ 2\beta_3) kV - \beta_3 kV^2]}{\gamma+(\beta_2 + 2\beta_3) kV - \beta_3 kV^2}.\label{eq:veq2}
\end{equation}
Linearizing about the $V=0$ equilibrium, we find that the epidemic threshold is given by the condition
\begin{equation}\label{eq:threshold_indiv_corr}
    \beta_2+2\beta_3={\gamma}\frac{\langle k \rangle}{\langle k^2 \rangle},
\end{equation}
which defines a linear relationship between $\beta_2$ and $\beta_3$ for fixed $\gamma$, in contrast to the collective contagion mechanism which does not alter the epidemic threshold $\beta_2^c=\gamma \langle k\rangle/\langle k^2\rangle$. This relationship can be understood heuristically by noting that, close to the $V=0$ solution, the probability that two nodes in a hyperedge are simultaneously infected can be neglected. Under that assumption, infection of a susceptible node by a triangle when at least one other node is infected is equivalent to independent infection by either of the two other nodes in the triangle with rate $\beta_3$. Since in the correlated case a node belongs, on average, to the same number of links and triangles, the individual contagion model reduces to the traditional SIS model with contagion rate $\beta_2^{\text{eff}} = \beta_2 + 2 \beta_3$ in the linear regime (we emphasize, however, that the nonlinear behavior can be different).

\begin{figure}[t]
    \centering
    \includegraphics[width=8cm]{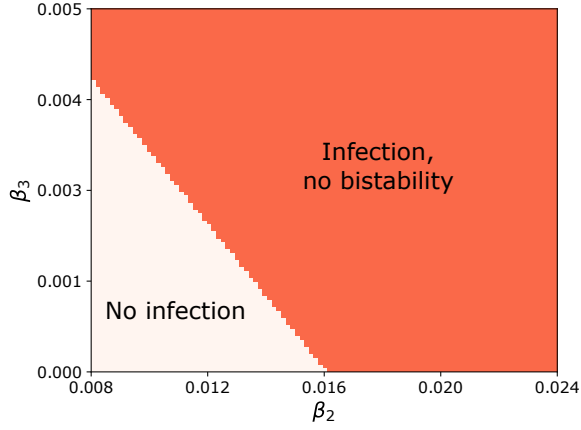}
    \caption{\label{fig:phase_space_individual}Phase diagram for the degree-correlated, individual contagion model with parameters $\gamma = 2$ and $P(k) \propto k^{-4}$ when $67 < k < 1000$ and $0$ otherwise.}
\end{figure}

In Fig.~\ref{fig:phase_space_individual} we plot the $(\beta_2,\beta_3)$ phase space for this scenario, with light pink indicating one solution ($V=0$) to Eq.~(\ref{eq:veq2}) and orange indicating two solutions, the unstable $V=0$ solution and a stable $V>0$ solution.

Considering the uncorrelated case where $f(k,k_1,k_2)=2\langle k\rangle/N^2$, and expressing Eq.~(\ref{eq:mfmarginal2}) in terms of $U$ and $V$, we obtain
\begin{eqnarray}
    \frac{dx_k}{dt}=&&-\gamma x_k+\beta_2(1-x_k) k V\\
    &&+2\beta_3(1-x_k)\langle k\rangle U - \beta_3(1-x_k)\langle k\rangle U^2\nonumber,
\end{eqnarray}
with equilibrium
\begin{equation}
    x_k=\frac{\beta_2 kV + 2\beta_3 \langle k\rangle U - \beta_3 \langle k\rangle U^2}{\gamma +\beta_2 kV + 2\beta_3\langle k\rangle U - \beta_3\langle k\rangle U^2},
\end{equation}
which has different first-order behavior than the degree correlated case. Inserting this expression into Eqs.~(\ref{eq:vdef}) and (\ref{eq:udef}), we obtain
\begin{align}
    U&=\frac{1}{N}\sum\limits_{k} \frac{P(k)(\beta_2 kV+ 2\beta_3 \langle k\rangle U - \beta_3 \langle k\rangle U^2)}{\gamma+\beta_2 kV + 2\beta_3 \langle k\rangle U - \beta_3 \langle k \rangle U^2},\\
    V&=\frac{1}{N\langle k\rangle}\sum\limits_{k} \frac{kP(k)(\beta_2 kV+ 2\beta_3 \langle k\rangle U - \beta_3 \langle k\rangle U^2)}{\gamma+\beta_2 kV + 2\beta_3 \langle k\rangle U - \beta_3 \langle k\rangle U^2}.
\end{align}
Linearizing, we obtain the system
\begin{align}
    \delta U&=\frac{\langle k\rangle \beta_2}{\gamma} \delta V+\frac{2\langle k\rangle \beta_3}{\gamma}\delta U,\\
    \delta V&=\frac{\langle k^2\rangle \beta_2}{\langle k\rangle\gamma} \delta V+\frac{2\langle k\rangle \beta_3}{\gamma}\delta U.
\end{align}
Solving this system and canceling the zero solution, we find that the epidemic threshold is defined by a non-linear relationship between the three epidemic parameters
\begin{equation}\label{eq:threshold_indiv_uncorr}
    \beta_2=\frac{\langle k\rangle \gamma^2 - 2\langle k\rangle^2\gamma \beta_3}{\langle k^2\rangle \gamma - 2(\langle k^2\rangle-\langle k\rangle^2)\langle k\rangle \beta_3}
\end{equation}

This relationship implies that there is a singularity when $\beta_3=\beta_3^{*}=\gamma\langle k^2\rangle/[2(\langle k^2\rangle -\langle k\rangle^2)\langle k\rangle]$. However, one can check that $\beta_2$ is negative at $\beta_3=\beta_3^{*}$, and therefore the singularity is not physically relevant. Note that when $\langle k^2\rangle = \langle k\rangle^2$ in the case of a $k$-regular network, the threshold reduces to that of the degree-correlated case.

\subsection{\label{sec:theory_higherorderhealing} Higher-order healing: hipster effect}

Here we consider the effect of higher-order healing for both collective and individual contagion. By \textit{higher-order healing} we refer to a situation where infected nodes that belong to a hyperedge of size $m>2$ with other infected nodes heal at rate $\beta_m$. This can be thought of as a ``hipster effect'' where if an idea or trend is popular in groups, then this makes an individual \textit{less} likely to adopt the trend, but the individual can be convinced to adopt the trend by their pairwise connections\cite{touboul2019hipster}. For both the collective and individual contagion cases, we comment on the existence of bistability based on numerical phase plots.

When the contagion is collective, the model including higher-order healing can be written as Eq.~(\ref{eq:mfmarginal}) with the sign of the third term changed, and because the triangle healing mechanism is solely higher-order, there is no effect on the epidemic threshold which is obtained by the linearization of the 0 solution. However, we find that explosive transitions do not occur for $\beta_2,\beta_3\geq 0$.

\begin{figure}
    \centering
    \includegraphics[width=8cm]{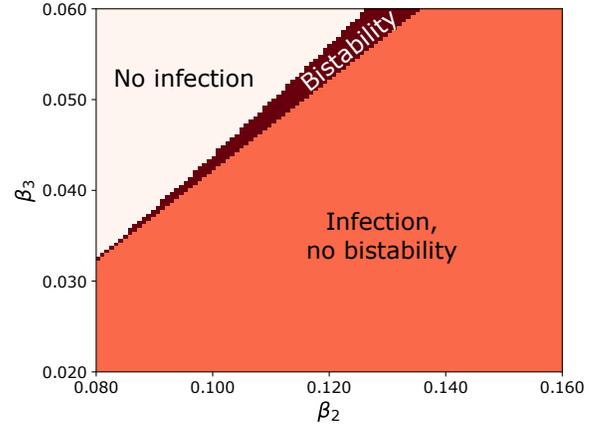}
    \caption{Phase diagram for the degree-correlated, higher-order healing with individual contagion with parameters $\gamma = 2$ and $P(k) \propto k^{-4}$ when $67 < k < 1000$ and $0$ otherwise.}
    \label{fig:phase_space_healing}
\end{figure}

Likewise, for the individual contagion model, higher-order healing can be written as Eq.~(\ref{eq:mfmarginal2}) with the third term negative. In this case, the epidemic threshold for both the degree-correlated and uncorrelated case can be obtained by substituting $-\beta_3$ for $\beta_3$ in Eq.~(\ref{eq:threshold_indiv_corr}) and Eq.~(\ref{eq:threshold_indiv_uncorr}) respectively. Higher-order healing in individual contagion enables explosive transitions to occur for ranges of $\beta_2,\beta_3\geq 0$, as can be seen in Fig.~\ref{fig:phase_space_healing} which shows the phase space $(\beta_2,\beta_3)$ for the degree-correlated case. As one might expect, for large enough higher-order healing $\beta_3$ there is no infection, but there is a narrow band of bistable behavior separating the regions of no infection and monostable infection.

\subsection{\label{sec:events_threshold}Unfortunate series of events}

So far we have considered hypergraphs with hyperedges of sizes 2 and 3 only. We now briefly discuss contagion in networks with hyperedges of all sizes, i.e. $M=N$. In the context of epidemic spreading, hyperedges could be interpreted as participation in social events such as parties, conferences, concerts, and sports events. For simplicity, we will focus on a hypergraph with degree-correlated hyperedges where
\begin{align}
f_m({\bf k},{\bf k}_1,\dots,{\bf k}_{m-1}) = \frac{(m-1)! \, k k_1 k_2 \dots k_{m-1}}{(N\langle k \rangle)^{m-1}} \frac{\langle k^{(m)}\rangle}{\langle k \rangle},
\end{align}
such that the average number of hyperedges of size $m$ a node belongs to is $\langle k^{(m)}\rangle$. In this case, by repeating the calculations of Sec.~\ref{sec:theory_individual}, the fraction of infected nodes of degree $k$ evolves in terms of the fraction of infected edges $V$ (\ref{eq:vdef}) as
\begin{equation}
    \frac{dx_k}{dt}=-\gamma +k(1-x_k)\sum\limits_{m=2}^M \frac{\beta_m \langle k^{(m)}\rangle}{\langle k\rangle}[1-(1-V)^{m-1}],\label{eq:series_unfortunate_individual}
\end{equation}
for individual contagion and
\begin{equation}
    \frac{dx_k}{dt}=-\gamma +k(1-x_k)\sum\limits_{m=2}^M \frac{\beta_m \langle k^{(m)}\rangle}{\langle k\rangle}V^{m-1},\label{eq:series_unfortunate_collective}
\end{equation}
for collective contagion. In the case of collective contagion, larger hyperedges can cause the emergence of new stable fixed points which can lead to richer consensus dynamics\cite{iacopini2019simplicial}. We focus, however, on the case of individual contagion. Linearizing, we find that the solution $x_k=0$ becomes unstable when
\begin{equation}
    \sum\limits_{m=2}^M\frac{(m-1)\beta_m \langle k^{(m)}\rangle}{\langle k\rangle} >\frac{\gamma \langle k\rangle}{\langle k^2\rangle}.\label{eq:series_condition}
\end{equation}

If the sum yields a value larger than ${\gamma}\langle k \rangle/\langle k^2 \rangle$ propagating social contagion will result. Social event restrictions implemented as a truncation of the series by prohibiting events larger than a certain size, or practices that reduce contagion in social events and reduce $\beta_m$  (such as enforcing physical separation) can reduce the value of the sum so that contagion does not propagate \cite{st2020school,althouse2020stochasticity}.

\section{\label{sec:critical_beta3}The effect of degree distribution on \texorpdfstring{$\beta_3^c$}{critical triangle infectivity}}

\begin{figure}[t]
\begin{center} 
\subfloat[Degree-correlated hyperedges]{\includegraphics[width=8cm]{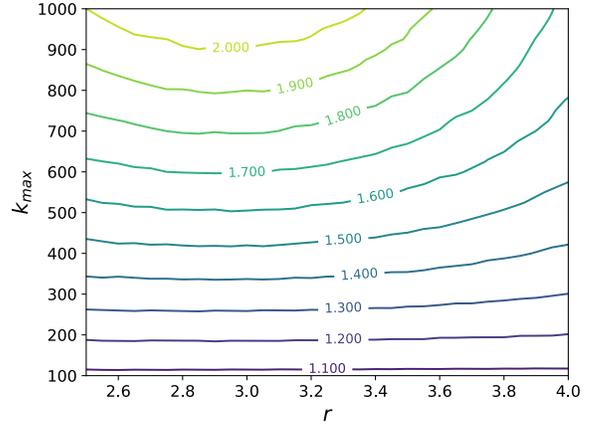}}\\ \subfloat[Uncorrelated hyperedges]{\includegraphics[width=8cm]{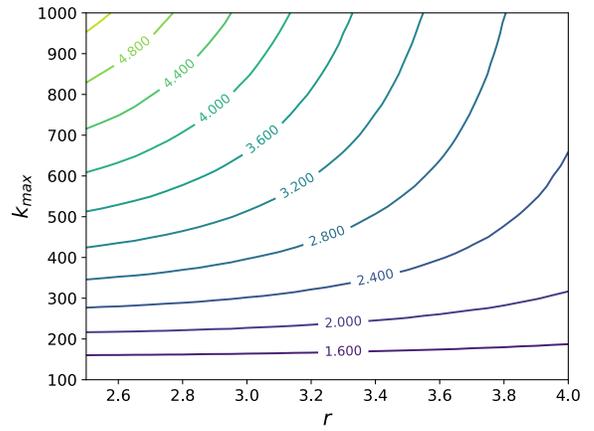}}\\
    \caption{$\beta_3^c/\beta_2^c$ as a function of power-law distribution parameters for the degree-correlated case (a) and the uncorrelated case (b). $\beta_3^c$ was calculated numerically from the mean field equations (see Appendix \ref{sec:appendix_beta3_critical}), and $\beta_2^c = \gamma \langle k \rangle/\langle k^2 \rangle$. The parameters are $P(k)\propto k^{-r}$ if $50 \leq k \leq k_{max}$ and $P(k)=0$ otherwise and $\gamma = 2$.}
    \label{fig:contour_critical_beta3}
\end{center}
\end{figure}

In Section \ref{sec:theory} we expressed the epidemic threshold $\beta_2^c$ in terms of moments of the degree distribution of the underlying network structure. Similarly, we would like to express the critical value of $\beta_3$ at which the explosive transitions appear, $\beta_3^c$, as a function of hypergraph structure. Explosive transitions and bistability occur when there are two stable steady-state solutions to Eqs.~(\ref{eq:mfmarginal}). For the degree-correlated and uncorrelated cases, this occurs when there are two non-zero solutions to Eq.~(\ref{eq:v1}) and the coupled system of Eqs.~(\ref{eq:ueq})-(\ref{eq:veq}) respectively. We can compute the critical value of $\beta_3$ by finding the numerical solution of these mean field equations and determining the value of $\beta_3^c$ at which bistability appears. This method is much more efficient than using stochastic microscopic simulations of the contagion model to infer the onset of explosive transitions and to map the phase space. Fig.~\ref{fig:contour_critical_beta3} shows the predicted value of $\beta_3^c$ normalized by $\beta_2^c$ for the correlated (a) and uncorrelated (b) cases as a function of the power-law exponent $r$ and the maximum degree $k_{max}$, where Eqs.~(\ref{eq:v1}) and (\ref{eq:ueq})-(\ref{eq:veq}) were solved using $P(k)\propto k^{-r}$ if $50 \leq k \leq k_{max}$ and $P(k)=0$ otherwise. Larger values of $r$ and $k_{max}$ correspond to larger heterogeneity of the degree distribution. We note that for the most homogeneous network -- the k-regular network -- $\beta_3^c/\beta_2^c$ is 1, and we see in Figs.~\ref{fig:contour_critical_beta3}(a) and (b) that $\beta_3^c$ increases relative to $\beta_2^c$ as $r$ or $k_{max}$ increase, except for small values of $r$ and large values of $k_{max}$ in the degree-correlated case. Thus, heterogeneity in the degree distribution of the pairwise interaction network appears to suppress explosive transitions. However, this effect is much more pronounced for the uncorrelated case (b) that for the degree-correlated case (a), as we discussed previously. In Appendix \ref{sec:appendix_beta3_critical}, we describe in more detail the algorithm employed to find $\beta_3^c$ from the mean field equations.

Although this method works well in predicting the value of $\beta_3^c$, it does not provide a direct relationship between the network structure and the onset of explosive transitions and is more computationally expensive than an analytical expression. For this reason, we present closed form approximations to $\beta_3^c$ and describe the parameter regimes over which they are accurate. Starting with the degree-correlated case and canceling the zero solution of Eq.~(\ref{eq:v1}), we find conditions under which there are at least two solutions to
\begin{equation}\label{eq:root_fcn_dc}
    h(V,\beta_2)=\frac{1}{N\langle k \rangle}\sum\limits_k \frac{k P(k)(\beta_2 k +\beta_3 k V)}{\gamma +\beta_2 k V +\beta_3 k V^2}-1=0.
\end{equation}

First, note that $h(0,\beta_2)=\beta_2/\beta_2^c-1$ and that $h(1,\beta_2)<0$. Therefore, if $\frac{\partial h}{\partial V}(0,\beta_2^c)>0$, then by continuity, there will be at least two solutions for $\beta_2$ less than, but sufficiently close to, $\beta_2^c$. This condition gives
\begin{equation}\label{eq:beta3_crit_corr}
    \frac{\beta_3^c}{\gamma}=\frac{\langle k^3\rangle\langle k\rangle^2}{\langle k^2\rangle^3},
\end{equation}
which works well in predicting the onset of bistability for the degree-correlated case. The relative error with respect to the value obtained from directly solving Eq.~(\ref{eq:v1}) for all distributions tested is less than 2\% (not shown).

The analysis for the degree-correlated case was based on the behavior of $h(V,\beta_2)$ near $V=0$. For the uncorrelated case, however, we find that a saddle-node bifurcation can occur at positive values of $V$, and it is necessary to expand Eqs.~(\ref{eq:ueq})-(\ref{eq:veq}) to higher order.


Expanding Eqs.~(\ref{eq:ueq})-(\ref{eq:veq}) to second order, setting $\beta_2=\beta_2^c=\gamma\langle k\rangle/\langle k^2\rangle$, and subtracting the two equations yields
\begin{equation}\label{eq:u_to_v}
    U=\frac{\langle k\rangle^2}{\langle k^2\rangle}V + \left(\frac{\langle k\rangle\langle k^3\rangle}{\langle k^2\rangle^2} -\frac{\langle k\rangle^2}{\langle k^2\rangle}\right)V^2,
\end{equation}
which, when evaluated in
\begin{equation}\label{eq:root_fcn_uc}
    h(V,\beta_2)=\frac{1}{N\langle k \rangle}\sum\limits_k \frac{k P(k)(\beta_2 kV +\beta_3 \langle k\rangle U^2)}{\gamma +\beta_2 k V +\beta_3 \langle k\rangle U^2}-V=0
\end{equation}
and expanded to fourth order, again setting $\beta_2=\beta_2^c$, yields
\begin{equation}
    h(V,\beta_2^c)=(a_0 + a_1 V + a_2 V^2)V^2,\label{eq:quadratic}
\end{equation}
where    
\begin{eqnarray}
    a_0&&=-\frac{\langle k\rangle\langle k^3\rangle}{\langle k^2\rangle^2}+\frac{\langle k\rangle^5\beta_3}{\langle k^2\rangle\gamma},\\
    a_1&&=\frac{\langle k\rangle^2\langle k^4\rangle}{\langle k^2\rangle^3} - 4\frac{\langle k\rangle^5 \beta_3}{\langle k^2\rangle^2\gamma} +2 \frac{\langle k\rangle^4\langle k^3\rangle\beta_3}{\langle k^2\rangle^3\gamma},\\
    a_2&&=-\frac{\langle k\rangle^3 \langle k^5\rangle}{\langle k^2\rangle^4}+5\frac{\langle k\rangle^5\beta_3}{\langle k^2\rangle^2\gamma}+3\frac{\langle k\rangle^6\langle k^3\rangle \beta_3}{\langle k^2\rangle^4\gamma}\\
    &&-6\frac{\langle k\rangle^4\langle k^3\rangle \beta_3}{\langle k^2\rangle^3\gamma}+\frac{\langle k\rangle^3 \langle k^3\rangle^2\beta_3}{\langle k^2\rangle^4\gamma}-\frac{\langle k\rangle^{10}\beta_3^2}{\langle k^2\rangle^4\gamma^2}.\nonumber
\end{eqnarray}

For continuous transitions to epidemics, there is only one equilibrium for $V$ at $\beta_2=\beta_2^c$, namely $V=0$. The onset of bistability occurs when a second solution appears, which corresponds to the first appearance of a root of (\ref{eq:quadratic}) in the interval $(0,1)$. Such a root can appear at $V=0$ in a transcritical bifurcation, or at $V>0$ as a pair of roots in a saddle-node bifurcation. A pair of roots appears when the discriminant of the quadratic equation $a_0+a_1V+a_2V^2=0$ is zero. However, this bifurcation is physically meaningless if it occurs for values of $V$ outside the interval $[0,1]$. Therefore, we impose the constraint that the value of $\beta_3$ found by solving $a_1^2-4a_0a_2=0$ must satisfy the inequality $0\leq -a_1/2a_2\leq 1$. In addition, we note that because of continuity, the sign of the $a_2$ term must be negative, because otherwise $\frac{\partial h}{\partial V}(0,\beta_2^c)>0$ and the bifurcation has already occurred. The transcritical bifurcation occurs when a root crosses from a negative value to a positive value, which occurs when one root of $a_0+a_1V+a_3V^2=0$ is $V=0$, implying that $a_0=0$ and $\beta_3^c=\gamma\langle k^3\rangle/\langle k\rangle^4$. Using these conditions, we can construct a piecewise definition of $\beta_3^c$
\begin{equation}
    \beta_3^c=\begin{cases} \text{Solve}(a_1^2-4\,a_0\,a_2 = 0),& a_2<0, 0\leq -\frac{a_1}{2a_2}\leq 1, \\ \frac{\langle k^3\rangle}{\langle k\rangle^4}\gamma,&\text{else}.\end{cases}\label{eq:cases}
\end{equation}

\begin{figure}
    \centering
    \includegraphics[width=8cm]{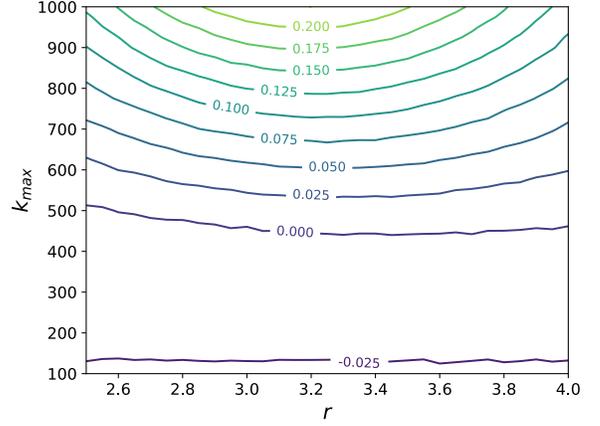}
    \caption{Relative error in the value of $\beta_3^c/\beta_2^c$ obtained from Eq.~(\ref{eq:cases}) compared with the numerically obtained value shown in Fig.~\ref{fig:contour_critical_beta3}(b).}
    \label{fig:higher_order_error_uncorrelated}
\end{figure}

The relative error in the value of $\beta_3^c/\beta_2^c$ obtained from Eq.~(\ref{eq:cases}) compared with the numerically obtained value shown in Fig.~\ref{fig:contour_critical_beta3}(b) is shown in Fig.~\ref{fig:higher_order_error_uncorrelated}.
In principle, one can expand to higher order to gain accuracy for the most heterogeneous of distributions. However, there is limited utility in increasing the order of the expansion further, because the resulting conditions become extremely complicated.

\section{\label{sec:discussion}Discussion}

In this paper we studied the SIS model of social contagion on hypergraphs with heterogeneous structure. The mean field description in Eq.~(\ref{eq:mf}) allowed us to explore the effects of hyperedge organization on the epidemic onset and the onset of bistability and explosive transitions.
One of our main findings is that with increasing heterogeneity of the pairwise network degree distribution, the onset of explosive transitions is postponed when the pairwise and higher order interactions have independent structure. More generally, when considering a hypergraph contagion model, the group infection and pairwise infection are competing mechanisms by which contagion spreads. Factors that promote contagion via pairwise infection, such as a heterogeneous degree distribution of the pairwise contact network, suppress discontinuous transitions. Conversely, heterogeneity in the degree distribution of hyperedges of higher order promotes such transitions. 

We considered two ways in which the structure of hyperedges of different sizes could be organized: the uncorrelated case, in which they are independent, and the correlated case, in which hyperedges of different sizes connect preferentially to the same nodes. While the organization of hyperedges in real world networks is surely much more complicated, these cases can be considered as null models against which the structure of real-world hypergraphs can be compared.

We studied various forms of higher-order contagion and healing: (i) collective contagion, in which all other members of the hyperedge need to be infected for contagion to occur, (ii) individual contagion, in which at least one member of the hyperdegree needs to be infected, and (iii) higher-order healing, in which pairwise interactions are infectious while higher-order interactions heal. Other forms of higher-order contagion could in principle be studied with the same methodology, but we leave these studies for future research.

Now we mention some of the limitations of our study. First, since we focused on the simplest contagion model, an important question left for future research is whether our results remain valid for more realistic epidemiological models (e.g. such as those used to model COVID-19 \cite{balcan2010modeling,arenas2020mathematical}). Our model also does not apply to non-Markovian contagion dynamics, which are important when modeling real-world epidemics. From a technical standpoint, another limitation is that we used a mean-field description of the dynamics, and it is known that such a description is not quantitatively accurate for moderate values of the infected population value \cite{newman2018networks,matamalas2020abrupt}. Since we were mainly interested in the behavior close to the onset of epidemics, the mean-field approximation was enough for our purposes. However, more precise descriptions could be obtained as in Refs.~\cite{matamalas2020abrupt,albert2002statistical}. Another important limitation of our hypergraph model is that we assume that the probability that two nodes belong to the same hyperedge is a function of their hyperdegrees. While this assumption can be relaxed by considering additional nodal variables, it is possible that such a model might be inadequate to describe some real-world networks. Finally we note that our model relies on knowledge of the functions $f_m$, which encode the organization of hyperedges across different hyperedge sizes. These functions have not yet been estimated from real-world networks, but as progress is made towards understanding the organization of higher-order interactions\cite{benson2018simplicial}, the determination of these functions could be a natural next step.

While in this paper we applied our hyperdegree based mean-field equation to the SIS epidemic model, the same formalism could be applied to other dynamical processes on hypergraphs, such as synchronization, opinion formation, and other types of epidemic models. We believe that this methodology will be useful to study the effect of heterogeneity on these hypergraph dynamical processes.
\vspace{0.25in}

\begin{acknowledgments}
Nicholas Landry acknowledges a helpful conversation with Danny Abrams and Llu\'is Arola-Fern\'andez on the ``hipster effect''.
\end{acknowledgments}

\nocite{*}

\appendix

\section{\label{sec:appendix_numerics}Numerical Experiments}

\subsection{\label{sec:appendix_microscopicsim}Microscopic simulation of the hypergraph SIS model}

We simulated the stochastic SIS model on a given hypergraph as a discrete-time Markov process on the nodes with transitions to infected and healthy states through the modalities described in Section \ref{sec:model_contagion}, a variant of which was simulated in Ref.~\cite{bodo2016sis}. We denote the binary states of the nodes at a time step $t$ by a vector ${\bf X}^t=(X_1^t,X_2^t,\dots,X_N^t)$ where $X_i^t=0$ if node $i$ is healthy and $X_i^t=1$ if it is infected. In this model, we assume that the events that a hyperedge infects node $i$ and that a pairwise connection infects node $i$ are independent. Likewise, we assume that an infected neighbor, whether through a pairwise or group connection, infects a node independently of any other neighboring node. The probability that a single infected node infects its pairwise neighbor in the time interval $[t,t+\Delta t]$ is $\beta_2\Delta t$, and so the probability that no neighboring node infects a given node is
\[(1-\beta_2\Delta t)^{(A{\bf X})_i}\]
where $A$ is the adjacency matrix with entries $A_{ij}=1$ if nodes $i$ and $j$ are connected by a link and 0 otherwise.

In the collective contagion model, the probability that a triangle infects a node in the time interval $[t,t+\Delta t]$ is $\beta_3\Delta t$ provided the other two nodes are infected. Therefore, the probability of no triangles infecting node $i$ can be written as
\[(1-\beta_3 \Delta t)^{\sum\limits_{\{i_1,i_2,i\}}X^t_{i_1}X^t_{i_2}},\]
where the sum is over all triangles $\{i_1,i_2,i\}$ with node $i$ as a member.
Lastly, the rate of healing is constant and independent of the infection status of any neighboring nodes so the probability that an infected node heals in a time interval $[t,t+\Delta t]$ is $\gamma \Delta t$.

The Markov process can then be described as
\begin{widetext}
\begin{align}
    P(X_i^{t+1}=1 \ | \ X_i^t=0) &= 1-(1-\beta_2 \Delta t)^{(A{\bf X^t})_i} (1-\beta_3 \Delta t)^{\sum\limits_{\{i_1,i_2,i\}}X^t_{i_1}X^t_{i_2}},\\
    P(X_i^{t+1}=0 \ | \ X_i^t=1) &= \gamma \Delta t.
\end{align}
\end{widetext}
In our simulations, we updated the status of the nodes synchronously at times $t=0,\Delta t,2\Delta t,\dots,n\Delta t$ where $\Delta t=0.1$.

Our specific implementation is described in what follows. We note that for all mechanisms of infection and healing described next, $u_i\sim\text{Uniform}(0,1)$ and this variable is drawn independently for each modality and each node $i$. At each time step, we iterate through every node and follow the following conditional logic. If a node $i$ is already infected, it is healed if $u_i<\gamma \Delta t$ and remains infected otherwise. Next, if the node $i$ is currently healthy, it is infected by its pairwise neighbors if $u_i<1-(1-\beta_2 \Delta t)^{(A{\bf X})_i}$ and remains healthy otherwise. If node $i$ still remains healthy after being subjected to pairwise infection, the node is infected by its triangle neighbors if
\[u_i<1-(1-\beta_3 \Delta t)^{\sum\limits_{\{i_1,i_2,i\}}X^t_{i_1}X^t_{i_2}}\]
and remains healthy otherwise. Note that each infection mechanism is only dependent on the prior time step so the order of these steps does not matter.

At $t=0$, the network is randomly and uniformly seeded with a small fraction ($p=0.001$) of infected nodes and at each subsequent step, the current state is iterated as described above and the population average, $x^t=\sum_{i=1}^N  X_i^t/N$ is stored. To avoid the absorbing state ${\bf X}^t={\bf 0}$, we infect a single randomly chosen node if the population becomes completely healthy. To mitigate the effect of variability in the stochastic simulation, we average the time response of $x^t$ over a sufficient time window (determined from the average infected response curves) after it reached the steady-state. In this study, we ran the simulation for a fixed set of parameters $\{\gamma,\beta_2,\beta_3\}$  for 1000 time steps and averaged over the last 300 time steps.

To find the bistability index, we initialize the simulation with a small fraction of infected agents for a fixed $\beta_3$ value and incrementally increase $\beta_2$ from a sufficiently small value (typically $\beta_2^c/2$) to a value above the critical value of $\beta_2$ (typically $3\beta_2^c/2$), and then incrementally decrease the value of $\beta_2$ down to its original value. As described previously, if the equilibrium value while increasing the value of $\beta_2$ is distinct from the equilibrium value while decreasing the value of $\beta_2$ for the same $\beta_2$ values, this indicates the presence of bistability. We simulated several equilibrium curves corresponding to different $\beta_3$ values to observe the value of $\beta_3$ at which the response curve starts to show bistability, and thus infer the value of $\beta_3^c$.

\subsection{\label{sec:appendix_networks}Network models} We exclusively considered networks generated using the configuration model in order to isolate the effect of the degree distribution. Although the configuration model has the potential to contain both self-loops and multi-edges, in practice, the fraction of these types of edges is small\cite{newman2018networks} and in our numerical experiments, the number of self-loops was approximately 1\% of the total number of nodes. 

We used networks of size $N=10^4$ in the simulation of the hypergraph SIS model because this was sufficiently large enough to reduce the finite-size effects. Because the network realization was relatively large, we did not average over an ensemble of these random graphs as in Ref.~\cite{iacopini2019simplicial}. We have described in Section \ref{sec:theory_2and3} the particular distributions examined.

We generated the triangles in two different ways corresponding to the two separate cases; degree-correlated and uncorrelated. For the first case, we used the same degree sequence as used to generate the network using the configuration model and extended the configuration model to triangles as has been done in prior work\cite{young2017construction,courtney2016generalized}. Because this is analogous to the construction of the network configuration model, there is also the possibility for self-loops and multi-edges, but this probability is low. For the independently-distributed triangles, we drew with replacement a fixed number of triples (enforcing the mean triangle degree) containing node indices and assigned these nodes to a triangle. Again, as with the standard configuration model, there is the possibility for self-loops and multi-edges, but the probability of either occurring is small.

\subsection{\label{sec:appendix_beta3_critical}The numerical computation of \texorpdfstring{$\beta_3^c$}{critical triangle infectivity}}

In Section \ref{sec:critical_beta3}, we plotted the numerical solution of $\beta_3^c$ for truncated power law distributions as a function of the maximum degree and power-law exponent. In this section, we discuss the specific methodology in generating these results.

First, we describe the process for finding the bistability index accurately from the mean-field equations (\ref{eq:v1}) and (\ref{eq:ueq})-(\ref{eq:veq}) for the correlated and uncorrelated cases respectively.
\begin{figure}
    \centering
    \includegraphics[width=8cm]{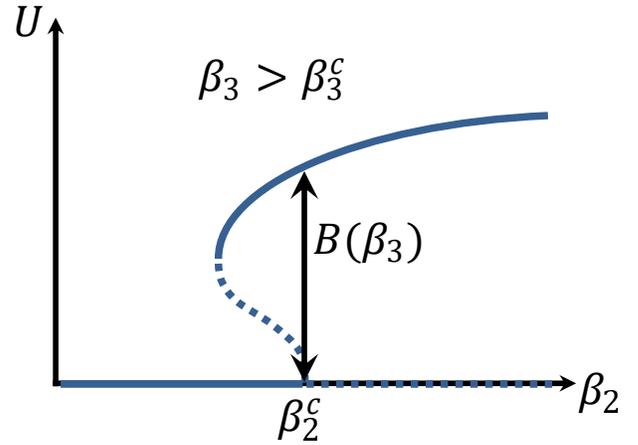}
    \caption{Illustration of the bistability index with respect to the solutions to the mean-field equation in the bistable regime}
    \label{fig:bistability_illustration}
\end{figure}
Since the $V=0$ solution becomes linearly unstable at $\beta_2=\beta_2^c$ and the stable $V>0$ solution is monotonically increasing with $\beta_2$, the bistability index $B(\beta_3)$ coincides with the value of the largest root of Eq. (\ref{eq:v1}) for the correlated case [or Eqs. (\ref{eq:ueq})-(\ref{eq:veq}) for the uncorrelated case] at $\beta_2=\beta_2^c$, as shown schematically in Fig. \ref{fig:bistability_illustration}. Therefore, using our analytical knowledge of $\beta_2^c$, we set $B(\beta_3)\approx V_{\epsilon}^*$, where $V_{\epsilon}^*$ is the largest root at $\beta_2=\beta_2^c-\epsilon$ with $\epsilon=10^{-5}$ a small number added for numerical robustness. (We verified that this method gives numerically accurate results when compared with other methods which do not require knowledge of $\beta_2^c$ but are more computationally intensive.)

Being able to compute $B(\beta_3)$, we find $\beta_3^c=\sup\{\beta_3 \ | \ B(\beta_3)=0\}$ by bisection: starting with an interval $[\beta_3^{min,0},\beta_3^{max,0}]$ such that $B(\beta_3^{min,0})=0$, $B(\beta_3^{max,0})>0$, we recursively define the interval $[\beta_3^{min,i+1},\beta_3^{max,i+1}]$ as $[\beta_3^{min,i},\widetilde{\beta}^i]$ if $B(\widetilde{\beta}^i)>0$ and $[\widetilde{\beta}^i,\beta_3^{max,i}]$ if $B(\widetilde{\beta}^i)=0$, where $\widetilde{\beta}^i=(\beta_3^{min,i}+\beta_3^{max,i})/2 $. When the length of the interval $[\beta_3^{min,i},\beta_3^{max,i}]$ is less that the tolerance $10^{-4}$, we set $\beta_3^c=\beta_3^{min,i}$.

\section*{Data Availability Statement} The data that support the findings of this study are available in Github at \href{https://github.com/nwlandry/SimplexSIS}{https://github.com/nwlandry/SimplexSIS}, Ref.~\cite{nicholas_landry_2020_4058981}.

\section*{References}

\bibliography{aip_heterogeneity_hypergraph_contagion}

\end{document}